\documentclass[11pt]{article}
\usepackage[utf8]{inputenc}
\usepackage{amsmath}
\usepackage{lipsum} 
\usepackage{fullpage}
\usepackage{mathrsfs}
\usepackage{svg}
\usepackage{graphicx}  
\graphicspath{{pictures/}}   
\usepackage{xr}
\usepackage{algorithm}
\usepackage{algpseudocode}
\usepackage{amsthm,cite,url,graphicx,booktabs,lipsum,color,bm,caption,subcaption,soul}
\usepackage{pifont,tikz,paralist,multirow,amssymb,float}
\usepackage{xifthen}
\usepackage{enumerate}
\usepackage{enumitem}
\usepackage{titlesec}
\usepackage{arydshln}
\usepackage[authoryear,round]{natbib}

\usepackage{color}
\usepackage{tikz}
\usepackage{multirow}

\newtheorem{theorem}{Theorem}

\newtheorem{proposition}{Proposition}

\usepackage{hyperref}
\hypersetup{
    colorlinks=true,
    linkcolor=blue,
    urlcolor=blue,
    citecolor=blue
}
\usetikzlibrary{positioning}
\usepackage{threeparttable}
 
\allowdisplaybreaks

\makeatother

\newcommand{\pr}{\mathbb{P}}

\newtheorem{innercustomgeneric}{\customgenericname}
\providecommand{\customgenericname}{}
\newcommand{\newcustomtheorem}[2]{%
  \newenvironment{#1}[1]
  {%
   \renewcommand\customgenericname{#2}%
   \renewcommand\theinnercustomgeneric{##1}%
   \innercustomgeneric
  }
  {\endinnercustomgeneric}
}
\newcustomtheorem{customthm}{Theorem}
\newcustomtheorem{customlemma}{Lemma}

\providecommand{\keywords}[1]{\textbf{\textit{Keywords:}} #1}

\def\E{\mathbb{E}}
\def\T{{\mathrm{\scriptscriptstyle T} }}

\def \E {\mathbb{E}}
\def\pr{\textnormal{pr}}

\makeatletter
\newcommand*{\indep}{%
 \mathbin{%
  \mathpalette{\@indep}{}%
 }%
}
\newcommand*{\nindep}{%
 \mathbin{
  \mathpalette{\@indep}{\not}
 }%
}
\newcommand*{\@indep}[2]{%
 \sbox0{$#1\perp\m@th$}
 \sbox2{$#1=$}
 \sbox4{$#1\vcenter{}$}
 \rlap{\copy0}
 \dimen@=\dimexpr\ht2-\ht4-.2pt\relax
 \kern\dimen@
 {#2}%
 \kern\dimen@
 \copy0 
}
\usepackage{xr}
\def\pr{\textnormal{pr}}

\def\T{{ \mathrm{\scriptscriptstyle T} }}
\makeatletter
\newcommand*{\addFileDependency}[1]{
  \typeout{(#1)}
  \@addtofilelist{#1}
  \IfFileExists{#1}{}{\typeout{No file #1.}}
}
\makeatother

 \usepackage{tocloft}
 \usepackage{setspace}

\begin{document}

\hypersetup{linkcolor=black}
\title{\bf A regression-based approach for bidirectional proximal causal inference in the presence of unmeasured confounding}  
\author{Jiaqi Min\textsuperscript{1}, Xueyue Zhang\textsuperscript{1},       Shanshan Luo\textsuperscript{1\footnote{Corresponding author: \href{mailto:shanshanluo@btbu.edu.cn}{shanshanluo@btbu.edu.cn}}}  \\\\	\textsuperscript{1} School of Mathematics and Statistics, \\Beijing Technology and Business University }

\date{} 
\maketitle  
\hypersetup{linkcolor=blue}
\begin{abstract}
Proxy variables are commonly used in causal inference when unmeasured confounding exists. While most existing proximal methods assume a
unidirectional causal relationship between two primary variables, many social
and biological systems exhibit complex feedback mechanisms that imply bidirectional
causality. In this paper, using regression-based models, we extend the proximal framework to identify bidirectional causal effects in the presence of unmeasured confounding. We establish the identification of bidirectional causal effects and develop a sensitivity analysis method for violations of the proxy structural conditions. Building on this identification result, we derive bidirectional two-stage least squares estimators that are consistent and asymptotically normal under standard regularity conditions. Simulation studies demonstrate that our approach delivers unbiased causal effect estimates and outperforms some standard methods. The simulation results also confirm the reliability of the sensitivity analysis procedure. Applying our methodology to a state-level panel dataset from 1985 to 2014 in the United States, we examine the bidirectional causal effects between abortion rates and murder rates. The analysis reveals a consistent negative effect of abortion rates on murder rates, while also detecting a potential reciprocal effect from murder rates to abortion rates that conventional unidirectional analyses have not considered.
\end{abstract}
\bigskip
\section{Introduction}   
Exploring causal relationships is essential for scientific research, especially in social, economic, and biological systems. These relationships are crucial for developing policies, estimating treatment effects, and evaluating interventions across various fields. For example, criminal justice research examines the effects of conviction versus incarceration on individual outcomes \citep{NBERw32894, kamat2024conviction}. In another domain, economic research investigates the association between tourism growth and economic development \citep{pulido2021analyzing}. However, identifying these relationships from observational data presents significant challenges because of unmeasured confounding in the exposure-outcome connection. A classic example of this problem is Simpson’s paradox \citep{simpson1951interpretation}, which shows how unmeasured confounding introduces bias. While direct adjustment methods such as inverse probability weighting, matching, regression, and doubly robust approaches can address observed confounding, most observational studies still face unmeasured confounding issues \citep{rubin1973use,rosenbaum1983central,stuart2010matching,bang2005doubly}. In such cases, causal effects cannot be uniquely identified from observational data without additional assumptions \citep{angrist1996identification,kang2016instrumental}. This may lead to biased results from traditional adjustment methods.

In recent years, there has been increasing interest in applying the proximal framework to address bias due to unmeasured confounding in observational data analysis  \citep{lipsitch2010negative,kuroki2014measurement,sofer2016negative,miao2018identifying,miao2024confounding,shi2020multiply,shi2020selective,Eric2023SS,cui2024semiparametric,luo2024causal}. Specifically, this approach utilizes three types of variables: (1) Observed confounders: variables that directly affect both the exposure and outcome; (2) Negative control exposures: variables that affect the treatment but are related to the outcome only through the unmeasured confounder; (3) Negative control outcomes: variables that affect the outcome but are linked to the treatment only through the unmeasured confounder. The variables in categories (2) and (3) are also commonly referred to as proxy variables or proxies in the literature. 

However, a key limitation of existing proximal causal inference methods is their assumption of unidirectional causal relationships. This assumption may be overly restrictive or unrealistic in many practical settings. For instance, \citet{wootton2021bidirectional} demonstrated reciprocal effects between loneliness and substance use.  \citet{carreras2018role} found that smoking reduces body mass index (BMI), and higher BMI increases the risk of smoking. When dealing with bidirectional relationships with unmeasured confounding, researchers have primarily relied on instrumental variable approaches, particularly two-sample Mendelian randomization  methods \citep{davey2014mendelian,bowden2015mendelian,xue2020inferring,hemani2017correction,li2024focusing}. Although several extensions have been proposed to handle invalid instruments \citep{xue2020inferring,darrous2021simultaneous}, they typically require strong assumptions that may not hold in practice and fail to leverage the rich information provided by proxy variables. The coexistence of bidirectional causality and unmeasured confounding poses a substantial methodological challenge in causal inference.

In this paper, we address these issues by extending the proximal causal inference framework to accommodate bidirectional relationships. Using regression-based models, we first show that bidirectional causal effects are identifiable when the conditional expectations of certain proxies satisfy specific nonlinear relationships. Building on this identification result, we introduce two bidirectional two-stage least squares (Bi-TSLS) estimators for each causal effect with corresponding estimation algorithms. We then establish their consistency and asymptotic normality under standard regularity conditions. Additionally, we develop a sensitivity analysis procedure to evaluate the performance under violations of the proxy structural conditions. Extensive simulation studies show that our approach outperforms classical methods across various scenarios. We apply our method to examine the bidirectional relationship between abortion rates and murder rates in the United States from 1985 to 2014. The results are consistent with prior findings, indicating that higher abortion rates are associated with lower murder rates. In addition, we identify a potential reverse effect, whereby increases in murder rates may be linked to decreases in abortion rates. Sensitivity analyses reveal that the estimated effect of abortion rates on murder rates remains stable across model specifications, while the reverse effect is more sensitive to violations of underlying assumptions.

The remainder of this paper is organized as follows. Section \ref{sec:methodology} proposes the bidirectional proximal causal model and establishes identification of causal effects in both directions, along with a sensitivity analysis method for violations of proxy structural conditions. Section \ref{sec:estimation} presents the Bi-TSLS estimators with their consistency and asymptotic properties. Section \ref{sec:simulation} investigates the finite sample performance through simulation studies and evaluates performance under various degrees of condition violations. Section \ref{sec:apply} applies the proposed methodology to explore bidirectional causality between abortion rates and murder rates, including sensitivity analysis to assess the stability of the findings. Section \ref{sec:conclusion} concludes with discussion of limitations and future research directions. Detailed proofs of theoretical results together with simulation and real data analysis codes, are provided in the online Supplementary Material. We also develop an \verb|R| package \verb|BiTSLS| that implements our proposed methodology, which is publicly accessible through both CRAN and GitHub \href{https://github.com/Jiaqi-Min/BiTSLS}{Jiaqi-Min/BiTSLS}.

\section{Methodology}\label{sec:methodology}
\subsection{Background and Preliminaries}

Throughout this paper, we assume there are $n$ individuals, independently and identically sampled from a superpopulation of interest. Let $X_i$ and $Y_i$ represent the treatment and outcome for individual $i$, which are used to explore causal relationships and effects. Let $V_i$ represent a vector of baseline covariates for individual $i$, and let $U_i$ denote the unmeasured confounder.  
In the proximal causal inference framework, two scalar variables, $Z_i$ and $W_i$, partially capture information about the unmeasured confounder $U_i$. Under this framework, $Z_i$ serves as a negative control exposure and $W_i$ as a  negative control outcome. We omit the subscript $i$ throughout the following discussion for notational simplicity. The variable relationships  within the proximal causal inference  framework are formally expressed as follows: 
\begin{align}
    \label{eq:nc-indep}
    W\indep (Z,X)\mid (U,V),~~Z \indep Y\mid (U,X,V).
\end{align}  
These conditions indicate that, after controlling for the unmeasured confounder $U$ and covariate $V$, the negative control outcome $W$ is independent of exposure $X$ and negative control exposure $Z$; similarly, after controlling for $U$, $X$, and $V$, the negative control exposure $Z$ is independent of outcome $Y$.  Fig.\ref{fig:causal-models}(a) provides a graphical illustration of this framework with covariate $V$ omitted for simplicity. A widely used linear structural model that satisfies   \eqref{eq:nc-indep} is:
\begin{align}
\label{eq:nc-linear}
\begin{gathered}
X = \alpha_0 + \alpha_z Z + \alpha_v V + \alpha_u U + \epsilon_x, \\
Y = \gamma_0 + \beta_{x \to y} X + \gamma_v V + \gamma_w W + \gamma_u U + \epsilon_y.
\end{gathered}
\end{align}
Here, the coefficient $\beta_{x \to y}$ represents the causal effect of $X$ on $Y$, which is typically the parameter of interest. The error terms $\epsilon_x$ and $\epsilon_y$ satisfy the moment restrictions $\mathbb{E}(\epsilon_x \mid Z, W, V, U) = 0$ and $\mathbb{E}(\epsilon_y \mid Z, W, V, U) = 0$, respectively.

Several recent studies have provided concrete examples of negative control variable selection to address unmeasured confounding.   \citet{shi2020multiply} investigated the effect of combination vaccines versus separate vaccines on fever risk among children, employing injury or trauma as a negative control outcome $W$ and ringworm as a negative control exposure $Z$ to mitigate potential confounding from unobserved health-seeking behaviors. In the social policy domain, \citet{mastouri2021proximal} and \citet{wu2024doubly} explored the relationship between abortion rates and murder rates in the United States. Given concerns about unobserved socioeconomic confounders, their analysis incorporated the generosity of Aid to Families with Dependent Children (AFDC) as a negative control exposure $Z$, while using prisoner population as a negative control outcome $W$.

\subsection{Identification under bidirectional causal model}\label{subsec:bidirection}
Although the proximal framework has been theoretically well developed in previous literature, many practical situations involve reciprocal causal relationships between variables. Specifically, a variable $X$ may influence $Y$ while $Y$ simultaneously affects $X$ through complex feedback mechanisms, as illustrated in Fig.\ref{fig:causal-models}(b). This bidirectional interaction is common in various fields, such as economics, healthcare, and social sciences. For example, pairs such as economic growth and inflation, or drug use and emotion regulation, often influence each other \citep{weiss2017testing,ramzi2019causality}.  
Such bidirectional causal structures pose substantial challenges for identification, particularly when proxy variables are used to account for unmeasured confounding. To characterize such bidirectional causal relationships, we extend the model \eqref{eq:nc-linear} to the following simultaneous linear structural equation model (SEM) \citep{li2024focusing}:
\begin{align}
\begin{gathered}
X=\alpha_0+\beta_{y\to x} Y +\alpha_v V+\alpha_zZ +\alpha_uU+\epsilon_x,\\
Y=\gamma_0+\beta_{x\to y} X +\gamma_v V+\gamma_wW   +\gamma_uU+\epsilon_y.
\end{gathered}\label{eq:bi-linear}
\end{align}  
In this bidirectional model, $\beta_{x\to y}$ represents the causal effect of $X$ on $Y$, while $\beta_{y\to x}$ represents the causal effect of $Y$ on $X$. Despite the presence of bidirectional causality between $X$ and $Y$, the model maintains certain proxy structural conditions: (i) the proxy variable $Z$ has a direct effect on the primary variable $X$ but does not directly affect $Y$;
(ii) the proxy variable $W$ has a direct effect on the primary variable $Y$ but does not directly affect $X$. We will later relax these structural conditions in Section \ref{subsec:sensitivity}, allowing both $W$ and $Z$ to have direct effects on   $X$ and $Y$. Model \eqref{eq:bi-linear} can be rewritten in the following reduced form \citep{hausman1983specification}: 
\begin{align}
\begin{aligned}
X&=\theta_0+ \theta_zZ +\theta_{w}W+\theta_v V +\theta_uU+\nu_x,\\
Y&=\mu_0+{\mu}_zZ+\mu_wW +\mu_v V +\mu_uU+\nu_y, 
\end{aligned}\label{eq:reduce}
\end{align}  
where $\iota=({1-\beta_{y\to x}\beta_{x\to y}})^{-1}$,  $\theta_0={(\alpha_0 +\gamma_0 \beta_{y\to x})\iota  }$, $\mu_0= {(\gamma_0 +\alpha_0 \beta_{x\to y})\iota  }$, $\theta_z=\alpha_z\iota $,  $\mu_z={\alpha_z\beta_{x\to y}}\iota $,   $\theta_w={\gamma_w }\beta_{y\to x}\iota $, $\mu_w={\gamma_w }\iota $,     $\theta_v={(\alpha_v +\gamma_v \beta_{y\to x})\iota  }$, $\mu_v= {(\gamma_v +\alpha_v \beta_{x\to y})\iota  }$,  $\theta_u={(\alpha_u +\gamma_u \beta_{y\to x})\iota  }$, $\mu_u= {(\gamma_u +\alpha_u \beta_{x\to y})\iota  }$, $\nu_x=({\epsilon_x+\epsilon_y\beta_{y \to x}})\iota$ and $\nu_y=(\epsilon_y+{\epsilon_x\beta_{x\to y}})\iota.$ From the reduced form, we can see that the bidirectional causal effects can be expressed in terms of the following coefficient ratios: 
\begin{equation}
    \label{eq:iv-est}
     \beta_{x \to y} =  {\mu_z }/ {\theta_z} ,~~  \beta_{y \to x} =   {\theta_w }/{ \mu_w}. 
\end{equation}
When both $\beta_{x \to y}$ and $\beta_{y \to x}$ are nonzero, model \eqref{eq:bi-linear} represents a cyclic graphical structure. The causal implications of such cyclic structures have been extensively studied in prior works \citep{lauritzen2002chain, hyttinen2012learning, rothenhausler2021anchor}. A common interpretation of cyclic SEM assumes that exogenous variables serve as background conditions that remain fixed until the system reaches equilibrium, a state referred to as deterministic equilibrium \citep{lauritzen2002chain}. As shown in Section S1 of the Supplementary Material, system stability requires that the spectral radius of the causal effect coefficient matrix be strictly less than 1. In consequence, we assume $|\beta_{x \to y}\beta_{y \to x}|<1$ in our bidirectional model. Under this condition, the cyclic relationships can be mathematically represented as systems that evolve to stable states, where our reduced-form equations in \eqref{eq:reduce} capture the final equilibrium relationships.

\begin{figure}[t]
\centering
\begin{tikzpicture}[>=latex, scale=0.9]
\begin{scope}[xshift=-5cm]
\node[draw, circle, thick] (Z) at (0,1.25) {$Z$};
\node[draw, circle, thick] (X) at (1.5,0) {$X$};
\node[draw, circle, thick] (Y) at (3.5,0) {$Y$};
\node[draw, circle, thick] (W) at (4.75,1.25) {$W$};
\node[draw, circle, thick] (U) at (2.5,1.25) {$U$}; 
\draw[thick, ->] (Z) -- (X);
\draw[thick, ->] (W) -- (Y);
\draw[thick, ->] (X) -- (Y); 
\draw[thick, ->] (U) -- (Z);
\draw[thick, ->] (U) -- (X);
\draw[thick, ->] (U) -- (Y);
\draw[thick, ->] (U) -- (W);
\node at (2.5,-1.5) {(a) Proximal causal model};
\end{scope}
\begin{scope}[xshift=5cm]
\node[draw, circle, thick] (Z) at (0,1.25) {$Z$};
\node[draw, circle, thick] (X) at (1.5,0) {$X$};
\node[draw, circle, thick] (Y) at (3.5,0) {$Y$};
\node[draw, circle, thick] (W) at (4.75,1.25) {$W$};
\node[draw, circle, thick] (U) at (2.5,1.25) {$U$};
\draw[thick, ->] (Z) -- (X);
\draw[thick, ->] (W) -- (Y);
\draw[thick, ->] (X) to[out=20,in=160] (Y);
\draw[thick, ->] (Y) to[out=200,in=340] (X);
\draw[thick, ->] (U) -- (Z);
\draw[thick, ->] (U) -- (X);
\draw[thick, ->] (U) -- (Y);
\draw[thick, ->] (U) -- (W);
\node at (2.5,-1.5) {(b) Bidirectional proximal causal model};
\end{scope}
\end{tikzpicture}
\caption{Graphical illustration of the standard and bidirectional proximal causal structures.}
\label{fig:causal-models}
\end{figure}

If the proxy variables $(Z,W)$ satisfy $Z \indep U$ and $W \indep U$, i.e., they serve as valid instrumental variables, the coefficients $\theta_z$, $\mu_z$, $\theta_w$, and $\mu_w$ in model \eqref{eq:reduce} can be directly identified and estimated using least squares \citep{xue2020inferring,li2024focusing}. Subsequently, the bidirectional causal effects in \eqref{eq:iv-est} can be straightforwardly obtained.
However, in practice, $Z$ and $W$ may be correlated with the unmeasured confounder $U$, in which case they are no longer valid instruments, and the coefficients $\theta_z$, $\mu_z$, $\theta_w$, and $\mu_w$ are no longer identifiable.  Moreover, when a bidirectional causal feedback mechanism exists between $X$ and $Y$, as described in \eqref{eq:bi-linear}, the standard proximal causal inference methods also fail to apply, since the conditional independence \eqref{eq:nc-indep}  is violated in this setting. 
To enable identification with these challenges, we introduce linear structural equation models for the proxy variables $Z$ and $W$, allowing them to be correlated with the unmeasured confounder $U$:
\begin{align}
Z=\delta_0+\delta_uU+\delta_vV +\epsilon_z,\label{eq:linear-ZW}~~W =\eta_0+\eta_uU +\eta_vV+\epsilon_w,
\end{align} 
where the two error terms satisfy $\E(\epsilon_z\mid U,V,W)=0$ and  $\E(\epsilon_w\mid U,V,Z)=0$.
\begin{theorem}\label{thm:iden}
Given models \eqref{eq:bi-linear} and \eqref{eq:linear-ZW}, 
we have (i) the causal parameter $\beta_{x\to y}$ is identifiable if $\E(W\mid V,Z)$ is a nonlinear function of $V$ and $Z$; (ii) the causal parameter $\beta_{y\to x}$ is also identifiable if $\E(Z\mid V,W)$ is a nonlinear function of $V$ and $W$.    
\end{theorem}
Theorem \ref{thm:iden} requires that  $\mathbb{E}(W \mid V, Z)$ be a nonlinear function of $V$ and $Z$. Combined with \eqref{eq:linear-ZW}, this implies that the conditional expectations $\mathbb{E}(\epsilon_z \mid V, Z)$ or $\mathbb{E}(U \mid V, Z)$  should  also be nonlinear functions of $V$ and $Z$. This assumption is usually satisfied in practice except in specific parametric model settings, such as when $(\epsilon_z, V, Z)$ are jointly normally distributed \citep{shuai2023identifying}. While Theorem~\ref{thm:iden} establishes identifiability under the condition that $\E(W \mid V, Z)$ and $\E(Z \mid V, W)$ are nonlinear functions, seemingly requiring correct model specification, we further show in Section~\ref{sec:estimation} that the proposed estimators remain consistent even when these working models are misspecified.

\subsection{Sensitivity analysis for model \eqref{eq:bi-linear}}\label{subsec:sensitivity}
In practical applications, the structural conditions in \eqref{eq:bi-linear} regarding proxy variables may not be fully satisfied, as $Z$ may directly affect $Y$, and $W$ may also directly affect $X$. This motivates a sensitivity analysis approach to evaluate how violations of these structural constraints influence causal effect estimation. To formalize this approach, we extend the structural equation model~\eqref{eq:bi-linear} as follows:
\begin{align}
\begin{aligned}
X & =\alpha_0+\beta^s_{y\to x} Y +\alpha_v V+\alpha_zZ 
 + \gamma_w R_w W +\alpha_uU+\epsilon_x,\\
Y &=\gamma_0+\beta^s_{x\to y} X +\gamma_v V+\gamma_wW  +\alpha_zR_zZ +\gamma_uU+\epsilon_y,
\end{aligned}\label{eq:bi-linear-sensitive}
\end{align}
where we introduce two sensitivity parameters, $R_w$ and $R_z$, to quantify the extent to which the structural conditions are violated. These parameters are equal to zero under model \eqref{eq:bi-linear}. This extension relaxes the original structural constraints by allowing both $Z$ and $W$ to directly affect both $X$ and $Y$ simultaneously. Increasing values of $|R_w|$ and $|R_z|$ indicate stronger direct effects of the proxy variables on the outcomes they were originally assumed not to influence, thereby reflecting a greater degree of violation of the proxy structural conditions. The corresponding reduced form maintains the same structure as in  \eqref{eq:reduce}, but with adjusted coefficients for $Z$ and $W$:
$\theta_z = (\alpha_z + \alpha_z R_z\beta^s_{y \to x})\iota$, $\theta_w = (\gamma_w R_w + \gamma_w\beta^s_{y \to x})\iota$, $\mu_z = (\alpha_z R_z + \alpha_z\beta^s_{x \to y})\iota$, and $\mu_w = (\gamma_w + \gamma_w R_w\beta^s_{x \to y})\iota$. All other coefficients retain the same form. 

Let  $S_{x\to y} = \mu_z/\theta_z$ and $S_{y\to x} = \theta_w/\mu_w$  denote the coefficient ratios, which share the same form as in equation~\eqref{eq:iv-est}.  As shown in Section S4 of the Supplementary Material, the  bidirectional causal effects with the given sensitivity parameters can be expressed as:
\begin{align}\label{eq:identify-sensitive}
\beta^s_{x\to y} = \frac{S_{x\to y}(1 + S_{y\to x}R_z - R_wR_z)-R_z}{1 - S_{x\to y}S_{y\to x}R_wR_z},~~
\beta^s_{y\to x} = \frac{S_{y\to x}(1 + S_{x\to y}R_w - R_wR_z)-R_w}{1 - S_{x\to y}S_{y\to x}R_wR_z}. 
\end{align}
The identification of bidirectional causal effects can be established under conditions similar to those in Theorem~\ref{thm:iden}, as stated in Proposition~\ref{prop:sensitivity}.
\begin{proposition}\label{prop:sensitivity}
Given models \eqref{eq:linear-ZW}  and \eqref{eq:bi-linear-sensitive} with sensitivity parameters $R_w$ and $R_z$, if ${1 - S_{x\to y}S_{y\to x}R_wR_z}\neq 0$,  
we have (i) the parameter $\beta^s_{x\to y}$ is identifiable if $\E(W\mid V,Z)$ is a nonlinear function of $V$ and $Z$; (ii) the parameter $\beta^s_{y\to x}$ is also identifiable if $\E(Z\mid V,W)$ is a nonlinear function of $V$ and $W$. 
\end{proposition}

\section{Estimation}\label{sec:estimation}
We now describe the estimation procedure for bidirectional causal effects. In principle, under models \eqref{eq:bi-linear} and \eqref{eq:linear-ZW},  it is possible to estimate the conditional expectation $\E(W \mid V, Z)$ and check whether it is linear,  although this may become complicated without a parametric specification for $\E(W \mid V, Z)$ \citep{Miao2021AJE}.  However, to estimate $\beta_{x\to y}$ and $\beta_{y\to x}$, the working models for these conditional expectations need not be correctly specified for the Bi-TSLS estimation. 

\begin{algorithm}
\caption{Bidirectional two-stage least squares  estimation for $\beta_{x\to y}$.}
\label{alg:betax2y}
\begin{algorithmic}[1]
\State   Fit a regression model $W \sim \delta_0 + \delta_zZ + \delta_{v1}V + \delta_{v2}m(Z, V)$ to obtain the fitted values $\hat{W} = \hat{\delta}_0 + \hat{\delta}_zZ + \hat{\delta}_{v1}V + \hat{\delta}_{v2}m(Z, V)$;

\State  Fit regression models $X \sim \theta_0 + \theta_zZ + \theta_w\hat{W} + \theta_vV$ and $Y \sim \mu_0 + \mu_zZ + \mu_w\hat{W} + \mu_vV$ to obtain the estimated coefficient  $\hat{\theta}_z$ and $\hat{\mu}_z$;

\State   Calculate the estimated causal effect: $\hat{\beta}_{x\to y} = {\hat{\mu}_z}/{\hat{\theta}_z}$.
\end{algorithmic}
\end{algorithm}
\begin{algorithm}
\caption{Bidirectional two-stage least squares estimation for $\beta_{y\to x}$.}\label{alg:betay2x}
\begin{algorithmic}[1]
\State   Fit a regression model $Z \sim \eta_0 + \eta_wW + \eta_{v1}V + \eta_{v2}h(W, V)$ to obtain the fitted values $\hat{Z} = \hat{\eta}_0 + \hat{\eta}_wW + \hat{\eta}_{v1}V + \hat{\eta}_{v2}h(W, V)$;

\State  Fit regression models $X \sim \theta_0 + \theta_z\hat{Z} + \theta_wW + \theta_vV$ and $Y \sim \mu_0 + \mu_z\hat{Z} + \mu_wW + \mu_vV$ to obtain the estimated coefficient $\hat{\theta}_w$ and $\hat{\mu}_w$;

\State   Calculate the estimated causal effect: $\hat{\beta}_{y\to x} =  {\hat{\theta}_w}/{\hat{\mu}_w}$.
\end{algorithmic}
\end{algorithm}

For clarity, let $m(Z, V)$ and $h(W, V)$ denote user-specified nonlinear functions, which may include interaction terms such as $ZV$ and $WV$ in practical applications. The detailed estimation procedures are outlined in Algorithms~\ref{alg:betax2y} and~\ref{alg:betay2x}, each comprising a two-stage process that incorporates the working models $m(Z, V)$ and $h(W, V)$. We then establish the consistency of the proposed estimators $\hat{\beta}_{x \to y}$ and $\hat{\beta}_{y \to x}$.

\begin{proposition} \label{prop:consistent}
Given models \eqref{eq:bi-linear} and \eqref{eq:linear-ZW}, we have (i) the proposed estimator  $\hat\beta_{x \to y}$  is consistent if $\delta_{v2} \neq 0$ and the  matrix   $  \E\left[\{1,Z, V, m(Z, V)\}^\T\{1,Z, V, m(Z, V)\}\right]$ has full rank;
(ii) the proposed estimator $\hat\beta_{y \to x}$  is also consistent if  $\eta_{v2} \neq 0$ and the matrix   $  \E\left[\{1,W, V, h(W, V)\}^\T\{1,W, V, h(W, V)\}\right]$ has full rank.
\end{proposition}

The full rank condition in Proposition \ref{prop:consistent} essentially requires the nonlinear condition in Theorem \ref{thm:iden}.  We formulate $\hat\beta_{x \to y}$ and $\hat{\beta}_{y \to x}$ as the solutions to a system of estimating equations (see the proof of Proposition \ref{prop:consistent} in Section S5 in the Supplementary Material). This formulation establishes the consistency of the estimator despite potentially misspecified working models $m(Z, V)$ and $h(W, V)$, while simultaneously providing the theoretical foundation for its asymptotic properties. 
Under standard regularity conditions, we establish the asymptotic normality of $\hat\beta_{x\to y}$ as stated in the following proposition.  Analogous results for $\hat\beta_{y\to x}$ can be established through similar arguments. 
\begin{proposition}
\label{prop:asymptotic} Given models \eqref{eq:bi-linear}, \eqref{eq:linear-ZW},    and the standard regularity conditions  in \citet{van2000asymptotic}, if $\delta_{v2} \neq 0$ and the matrix $\E\left[\{1,Z, V, m(Z, V)\}^\T\{1,Z, V, m(Z, V)\}\right]$ has full rank,  $n^{1/2}(\hat\beta_{x\to y}- \beta_{x\to y})$ converges in distribution to $N(0,\Sigma)$ as $n \to \infty$, where   the detailed form of $\Sigma$ is provided in Section S6 of the Supplementary Material.  
\end{proposition}
To conclude this section, we provide some remarks on estimation procedures in the context of sensitivity analysis.  
As shown in equation~\eqref{eq:identify-sensitive}, the  bidirectional causal effects $\beta^s_{x \to y}$ and $\beta^s_{y \to x}$ still depends on the quantities $S_{x \to y}$ and $S_{y \to x}$. These parameters  can also  be estimated using Algorithms~\ref{alg:betax2y} and~\ref{alg:betay2x}, respectively, following the same estimation procedures, resulting in two consistent estimators $\hat{S}_{x \to y}$ and $\hat{S}_{y \to x}$. Subsequently, these estimators can be plugged into equation \eqref{eq:identify-sensitive} to derive the  estimators $\hat{\beta}^s_{x\to y}$ and $\hat{\beta}^s_{y\to x}$. Treating $R_w$ and $R_z$ as fixed known constants, $\hat{\beta}^s_{x\to y}$ and $\hat{\beta}^s_{y\to x}$ are consistent and asymptotically normal under similar conditions as $\hat{\beta}_{x\to y}$ and $\hat{\beta}_{y\to x}$. The complete estimation procedures and proof of consistency and asymptotic normality are provided in Section S7 in the Supplementary Material.

\begin{figure}[ht]
    \centering
    \includegraphics[width=0.9\linewidth]{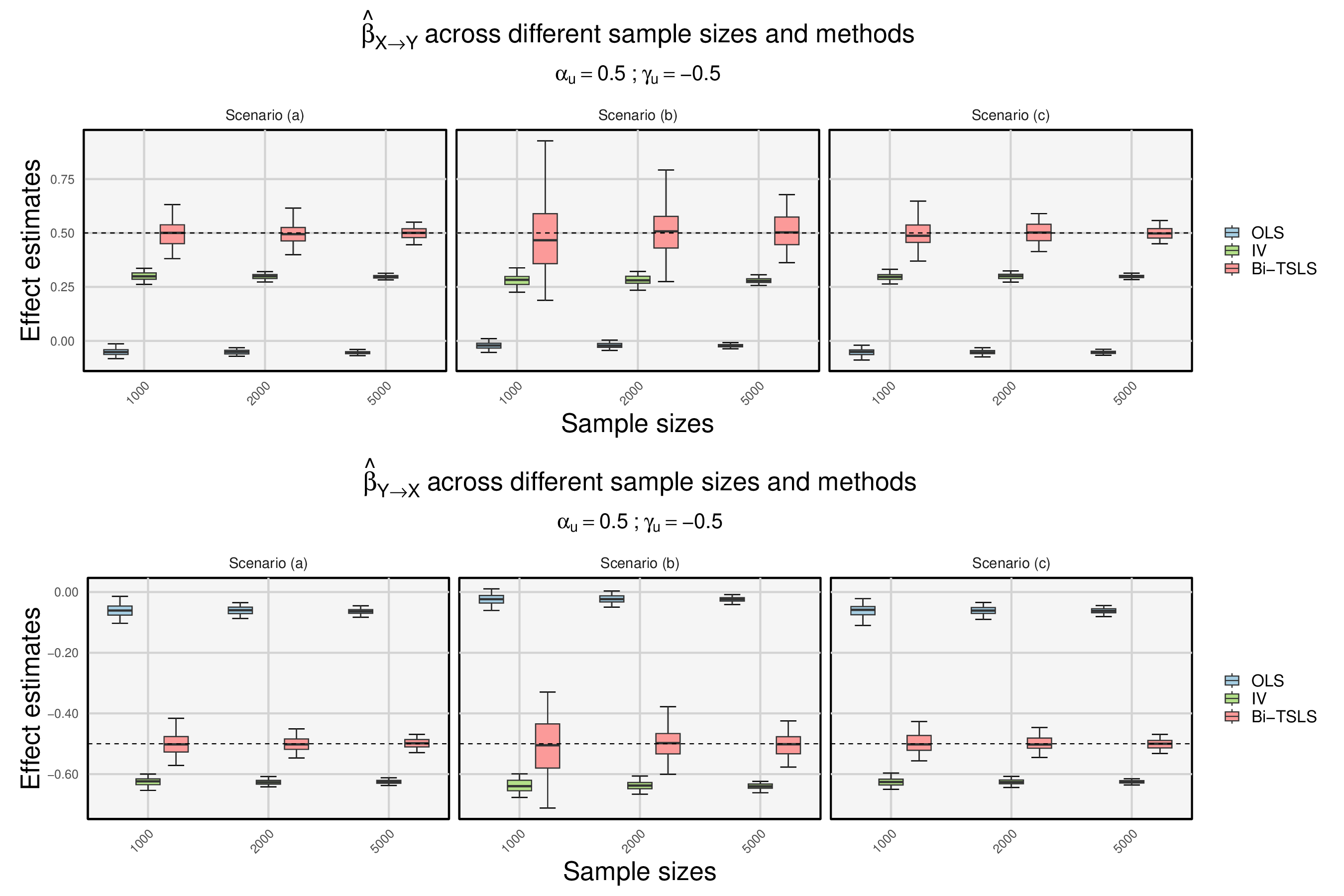}
    \caption{Estimates for bidirectional causal effects across different methods, sample sizes, and scenarios.}
    \label{fig:combined_effects}
\end{figure}

\section{Simulation Experiments}\label{sec:simulation}
\subsection{Simulation Settings}\label{subsec:performance}
In this section, we evaluate finite-sample performance of the proposed method. The performance of the theoretical framework is evaluated through simulation studies with the following data generation process:
\begin{enumerate}
\item Generate the observed covariate $V$  from a standard normal distribution, $V\sim N(0,1)$.
\item Generate the unmeasured confounder by setting $U=\mathrm{exp}(V)+ \epsilon_u$, 
where $ \epsilon_u$ is generated from the Rademacher distribution with probabilities $\pr(\epsilon_u =1)=\pr(\epsilon_u =-1)=0.5.$ 
\item Generate the proxy variables  $Z$ and $W$ from   models \eqref{eq:linear-ZW}, where $\delta_0=1, \delta_u=1,\delta_v=-0.5, \eta_0=1,\eta_u=-1,\eta_v=0.5$. We consider the following three scenarios for noise terms $\epsilon_z$ and $\epsilon_w$:
\begin{itemize}
\item[(a)]  $\epsilon_z\sim N(0,1)$ and  $\epsilon_w\sim N(0,1)$,
\item[(b)]   $\epsilon_z\sim \mathrm{Unif}[-1,1]$ and  $\epsilon_w\sim  \mathrm{Unif}[-1,1]$,
\item[(c)]  $\epsilon_z $ and  $\epsilon_w $ are generated from $\pr(\epsilon_z =1)=\pr(\epsilon_z =-1)=\pr(\epsilon_w =1)=\pr(\epsilon_w =-1)=0.5.$
\end{itemize}
In the simulation studies, we examine three noise settings for the proxy variables to evaluate the robustness of our identification strategy in Theorem \ref{thm:iden} under varying levels of nonlinearity. 
\item Generate the primary variables $(X, Y)$, where $X = \tilde{X}_{5000}$, $Y = \tilde{Y}_{5000}$, and the full time series $\{\tilde{X}_t, \tilde{Y}_t\}_{t=1}^{5000}$, using the following bidirectional iterative process:
\begin{equation}
\begin{aligned}
\tilde{X}_t &= \alpha_0 + \beta_{y \to x} \tilde{Y}_{t-1} + \alpha_v V + \alpha_z Z + \alpha_u U + \epsilon_x, \\
\tilde{Y}_t &= \gamma_0 + \beta_{x \to y} \tilde{X}_{t-1} + \gamma_v V + \gamma_w W + \gamma_u U + \epsilon_y,
\end{aligned}
\label{eq:iteration}
\end{equation}
where the parameters are set as follows: $\alpha_0 = 1$, $\beta_{y \to x} = -0.5$, $\alpha_v = 1$, $\alpha_z = 1$, $\alpha_u = 0.5$, $\gamma_0 = -1$, $\beta_{x \to y} = 0.5$, $\gamma_v = -1$, $\gamma_w = 2$, and $\gamma_u = -0.5$. The error terms $(\epsilon_x, \epsilon_y)$ follow independent Rademacher distributions, taking values $+1$ or $-1$ with equal probability 0.5. This bidirectional iterative process \eqref{eq:iteration} reflects the dynamics of reciprocal causal effects and converges to a stationary equilibrium characterized by the reduced-form system in  \eqref{eq:reduce}. More details are provided in Section S2 of the Supplementary Material.
\end{enumerate}

\subsection{Estimation Results}
 For each scenario, we conducted 200 simulation replications with sample sizes of $n =$ 1000, 2000, and 5000. We compare three estimation methods: (i) our proposed Bi-TSLS estimation, (ii) ordinary least squares (OLS) regression estimation, and (iii) instrumental variable (IV) estimation. For the Bi-TSLS estimation, we specified $m(Z, V)=ZV$ in Algorithm \ref{alg:betax2y} and $h(W, V) = WV$ in Algorithm \ref{alg:betay2x}. For OLS, we obtain $\hat\beta_{x\to y}^{\mathrm{ols}}$ and $\hat\beta_{y\to x}^{\mathrm{ols}}$ by fitting the regressions:
$$ X \sim \beta_{y\to x}^{\mathrm{ols}} Y + \theta_z^{\mathrm{ols}} Z + \theta_w^{\mathrm{ols}} W + \theta_v^{\mathrm{ols}} V,~~ Y \sim \beta_{x\to y}^{\mathrm{ols}} X + {\mu}_z^{\mathrm{ols}} Z + \mu_w^{\mathrm{ols}} W + \mu_v^{\mathrm{ols}} V.$$ 
For IV estimation, we obtain $ \hat\beta_{y\to x}^{\mathrm{iv}} = \hat\theta_{w}^{\mathrm{iv}} / \hat\mu_{w}^{\mathrm{iv}} $ and $ \hat\beta_{x\to y}^{\mathrm{iv}} = \hat{\mu}_z^{\mathrm{iv}} / \hat\theta_z^{\mathrm{iv}} $ by fitting the regressions: $$ X \sim \theta_z^{\mathrm{iv}} Z + \theta_w^{\mathrm{iv}} W + \theta_v^{\mathrm{iv}} V , Y \sim {\mu}_z^{\mathrm{iv}} Z + \mu_w^{\mathrm{iv}} W + \mu_v^{\mathrm{iv}} V.$$  

Fig.\ref{fig:combined_effects} presents boxplots of the three estimators across all scenarios and sample sizes with dashed lines indicating the true values $ \beta_{x\to y}= 0.5$ and $\beta_{y\to x} =-0.5$, respectively.  Due to the presence of unmeasured confounding, we find that OLS estimates are biased in all settings. Furthermore, since the two proxy variables $Z$ and $W$ cannot be considered valid instrumental variables, both $\hat{\beta}_{x \to y}^{\mathrm{iv}}$ and $\hat{\beta}_{y \to x}^{\mathrm{iv}}$  exhibit substantial bias.  In contrast, our proposed Bi-TSLS estimators perform well, providing consistent estimates across all scenarios, with the estimation variance decreasing as the sample size increases.

\begin{figure}[ht]
    \centering
    \includegraphics[width=0.9\linewidth]{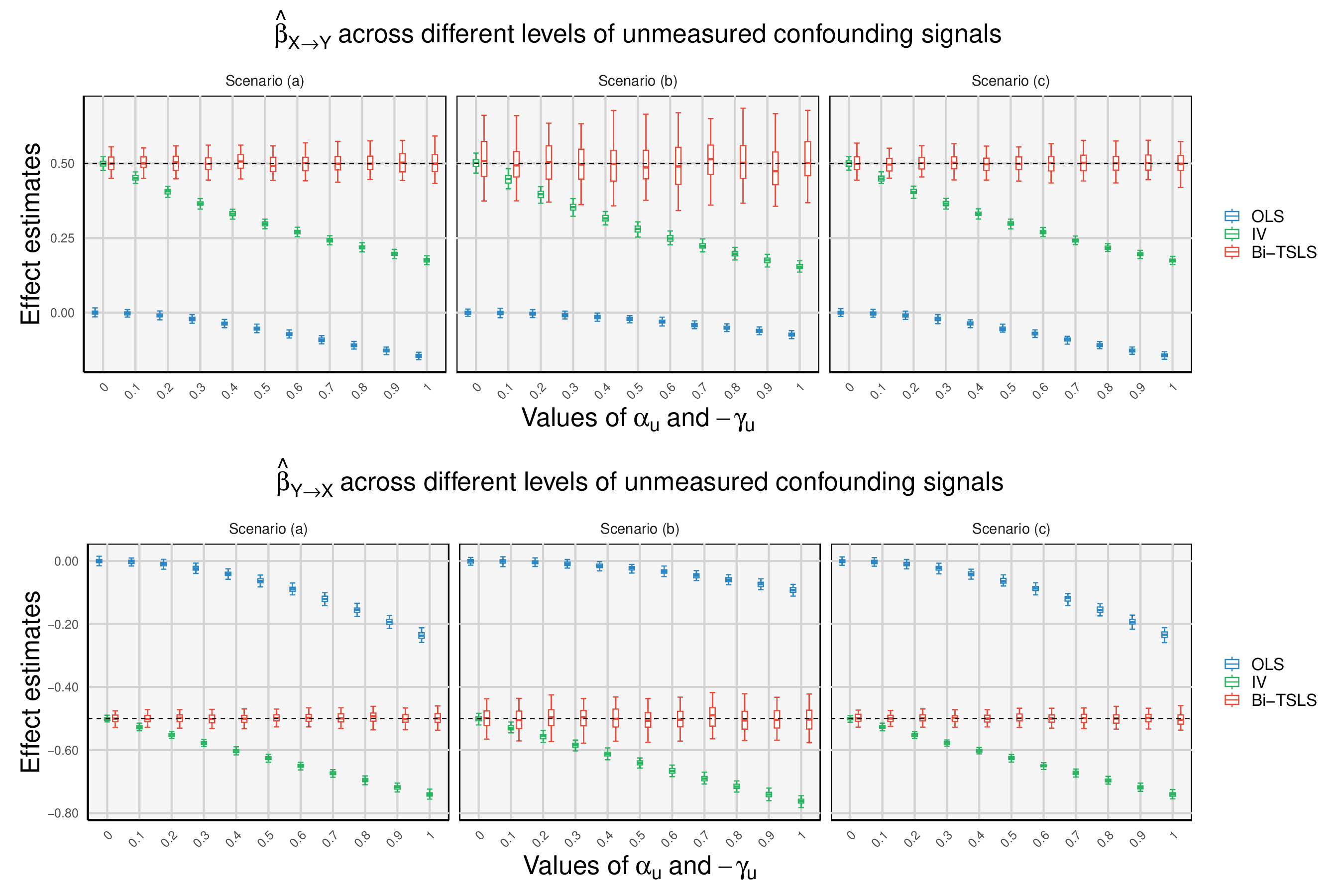}
    \caption{Estimates for bidirectional causal effects across different levels of unmeasured confounding signals.}
    \label{fig:strength}
\end{figure}

We also examine how varying the unmeasured confounding coefficients $\alpha_{u}$ and $\gamma_{u}$ in \eqref{eq:iteration} affects our estimator's performance, which represent different levels of confounding signals. For simplicity, the estimation results are based on a sample size of 5000. As illustrated in Fig.\ref{fig:strength}, although the proposed method exhibits a slight increase in estimation  variance with larger $|\alpha_{u}|$ and $|\gamma_{u}|$, the Bi-TSLS estimators maintain unbiased and stable variance across different levels of confounding signals. These desirable properties are confirmed in all three scenarios and for both bidirectional causal effects, $\beta_{x\to y}$ and $\beta_{y \to x}$. The IV estimators gradually approach the true values as $|\alpha_u|$ and $|\gamma_u|$ decrease, achieving unbiased estimation only when unmeasured confounding is absent. The OLS estimators still demonstrate bias across all levels of confounding signals.

\subsection{Simulation for sensitivity analysis}
We evaluate the sensitivity analysis performance when the model \eqref{eq:bi-linear} is violated by modifying the data generation process \eqref{eq:iteration} by allowing $W$ to influence $X$ directly through the coefficient $\gamma_w R_w$ and $Z$ to directly affect $Y$ through the coefficient $\alpha_z R_z$:
\begin{equation}
\begin{aligned}
\tilde{X}_t &= \alpha_0 + \beta^s_{y \to x}\tilde{Y}_{t-1} + \alpha_vV + \alpha_zZ + \gamma_w R_wW + \alpha_uU + \epsilon_x, \\
\tilde{Y}_t &= \gamma_0 + \beta^s_{x \to y}\tilde{X}_{t-1} + \gamma_vV  + \gamma_wW + \alpha_z R_zZ + \gamma_uU + \epsilon_y.
\end{aligned}
\end{equation}
We set the true values $\beta^s_{x\to y} = 0.5$ and $\beta^s_{y\to x} = -0.5$, consistent with the main simulation, and keep all other parameter settings the same as in Section \ref{subsec:performance}. In addition, we vary the sensitivity parameters $R_w$ and $R_z$ over the range [-0.5, 0.5] in increments of 0.1 to examine the performance of our sensitivity-adjusted estimators under different levels of proxy structural conditions violation.

\begin{figure}[ht]
    \centering
    \includegraphics[width=0.9\linewidth]{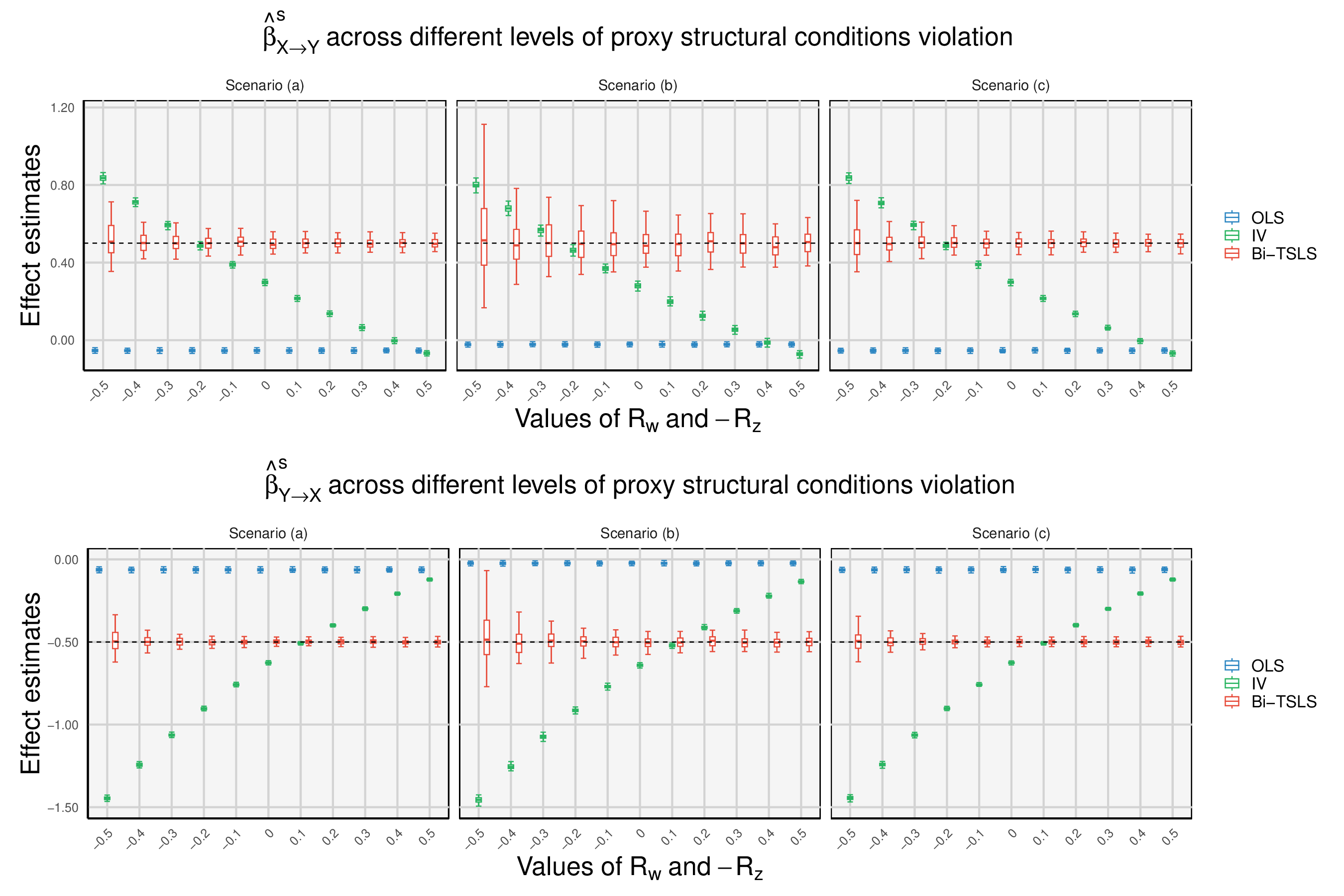}
    \caption{Estimates for bidirectional causal effects across different levels of proxy structural conditions violation.}
    \label{fig:sensitive}
\end{figure}

Fig.\ref{fig:sensitive} illustrates the estimation results based on a sample size of 5000 for simplicity. It can be observed that the proposed estimators accurately recover the true causal effects when the correct sensitivity parameters are used in the adjustment formulas \eqref{eq:identify-sensitive}. In contrast to our adjusted Bi-TSLS estimators, the IV estimators exhibit bias that varies dramatically as $R_w$ and $-R_z$ move from negative to positive values. The OLS estimators exhibit persistent bias across all settings.

\section{Real Data Analysis}\label{sec:apply}
\subsection{Description and estimation results within model \eqref{eq:bi-linear}}
In this section, to illustrate the practical utility of our methodology, we analyze state-level panel data from 1985 to 2014 to investigate the bidirectional causal relationship between abortion rates and murder rates in the United States. This topic has generated debate in social science research since \citet{donohue2001impact} proposed the hypothesis that increased access to abortion contributed to the substantial decline in murder rates observed in subsequent decades. They argued that unwanted children may face adverse circumstances that increase their likelihood of engaging in criminal behavior. Consequently, increased abortion access may lead to lower murder rates. While some studies have examined this unidirectional relationship \citep{donohue2001impact,donohue2020impact,woody2020estimating,mastouri2021proximal,wu2024doubly}, the possibility of a reciprocal effect, where murder rates might influence abortion decisions through some socioeconomic pathways, remains largely unexplored.

Our analytical sample includes annual observations for 48 states, yielding a panel with 1,440 state-year observations. We treat the abortion rate as the primary variable $X$ and denote the logarithm of the murder rate as $Y$. Following \citet{woody2020estimating}, we include baseline covariates $V$ comprising census population data, the state unemployment rate, the log of state per capita income, and the state poverty rate. We use the generosity of Aid to Families
with Dependent Children (AFDC) as the negative control exposure $Z$. Welfare generosity shapes fertility and therefore abortion decisions, yet once socioeconomic channels are adjusted for, it does not directly affect contemporaneous homicide. 
For the negative control outcome $W$, we choose the log of prisoner population per capita, as incarceration rates reflect violent crime levels and enforcement priorities but may not influence abortion rates \citep{donohue2020impact}. Moreover, findings from \citet{woody2020estimating} indicate that AFDC ($Z$) primarily influences abortion rates ($X$) while prison population ($W$) operates solely on the murder rates ($Y$), supporting our proxy choices. This selection of proxy variables aligns with previous unidirectional proximal causal inference research \citep{mastouri2021proximal,wu2024doubly}.

We first compare the results obtained by our Bi-TSLS approach and existing estimation methods. Table~\ref{tab:all state} summarizes point estimates, standard errors (SE), and 95\% confidence intervals (95\% CI) based on 200 bootstrap samples for OLS, standard IV, and our Bi-TSLS estimator. For the effect of abortion rates on murder rates, our Bi-TSLS estimate indicates a significant negative effect $\hat{\beta}_{x\to y}=-0.617$ (95\% CI: (-0.760, -0.497)), consistent with prior findings. In the reverse direction, we find that higher murder rates are associated with lower abortion rates ($\hat{\beta}_{y\to x}= -0.313$), indicating a significant effect of $Y$ on $X$ (95\% CI: (-0.452, -0.171)). However, while IV estimator suggests a slight negative effect ($\hat{\beta}^{\mathrm{iv}}_{y\to x}= -0.003$), the effect is not statistically significant (95\% CI: (-0.115, 0.109)). The OLS estimates show bidirectional negative associations ($\hat{\beta}^\mathrm{ols}_{x\to y}=-0.238$, $\hat{\beta}^\mathrm{ols}_{y\to x}=-0.390$), though these are likely biased due to unmeasured confounding. 

These findings suggest a complex bidirectional relationship between abortion rates and murder rates. The negative effect of abortion rates on murder rates supports the hypothesis that reduced unwanted births may contribute to lower future criminal behavior. The reverse causation indicates that high-crime environments may influence reproductive decisions, potentially through economic constraints or reduced healthcare accessibility. However, the robustness of these estimates requires further examination through sensitivity analyses to assess their stability under various conditions.
 
\begin{table}[ht]
\centering
\resizebox{0.7\textwidth}{!}{%
\begin{tabular}{cccccccc}
\toprule
\multirow{2}{*}{Method} & \multicolumn{3}{c}{Causal effect ${X\to Y}$} &  & \multicolumn{3}{c}{Causal effect ${Y\to X}$} \\ \cline{2-4} \cline{6-8} \addlinespace[1mm]
& Estimate        & SE       & 95\%CI       &  & Estimate        & SE       & 95\%CI       \\ \hline\addlinespace[1mm]
OLS & -0.238 & 0.001 & (-0.270, -0.206) &  & -0.390 & 0.001 & (-0.446, -0.336) \\
IV & -0.252 & 0.002 & (-0.347, -0.163) &  & -0.003 & 0.003 & (-0.115, 0.109) \\
Bi-TSLS & -0.617 & 0.003 & (-0.760, -0.497) &  & -0.313 & 0.004 & (-0.452, -0.171) \\
\bottomrule
\end{tabular}%
}
\caption{Estimated bidirectional causal effects between abortion rates and murder rates using different methods.}
\label{tab:all state}
\end{table}
\begin{figure}[ht]
    \centering
    \includegraphics[width=0.95\linewidth]{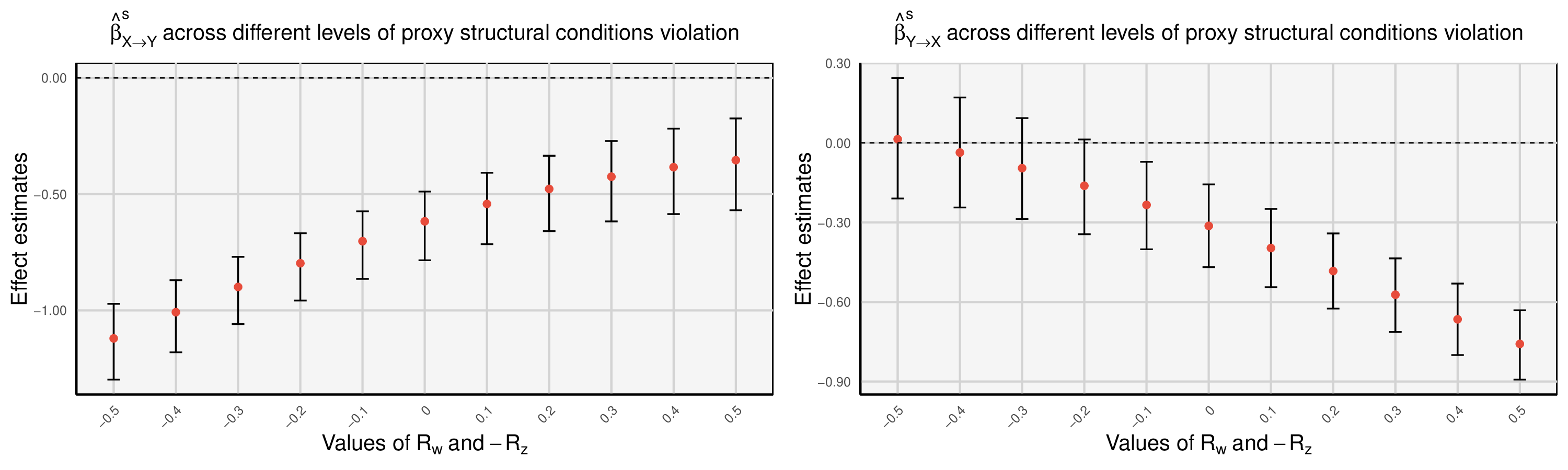}
    \caption{Sensitivity analyses across different levels of proxy structural conditions violation.}
    \label{fig:data}
\end{figure}

\subsection{Sensitivity analysis estimation results}
We examine the robustness of our estimated bidirectional causal effects by conducting sensitivity analysis with respect to violations of the proxy structural conditions. 
Fig.\ref{fig:data} provides the point estimates and 95\% confidence intervals, across a range of values for $R_w$ and $-R_z$ from $-0.5$ to $0.5$. The left panel presents the estimated effect of abortion rates on murder rates ($\hat{\beta}^s_{x \to y}$). The right panel shows the estimated effect of murder rates on abortion rates ($\hat{\beta}^s_{y \to x}$).

{Sensitivity analysis reveals notable differences in the robustness of the two causal directions.  The causal effect of abortion rates on murder rates remains consistently negative and statistically significant across all values of the sensitivity parameters, indicating that this effect is robust under various model specifications. Although the magnitude diminishes as $R_w$ and $-R_z$ increase, the direction of the effect does not change.
In contrast, the causal effect of murder rates on abortion rates is more sensitive to violations of the proxy structural conditions. While the estimated effect is generally negative, it crosses zero when $R_w$ and $-R_z$ fall below approximately $-0.2$, and the confidence intervals include zero under certain violations. This suggests that the reverse causal effect may be less robust to model misspecification compared to the primary effect. These findings highlight the importance of carefully assessing assumption violations when interpreting bidirectional causal relationships.}

\section{Conclusion}\label{sec:conclusion}
This paper extends proximal causal inference to settings involving bidirectional causal relationships with unmeasured confounding. We establish the identifiability of bidirectional causal effects by leveraging nonlinear proxy information and propose  two Bi-TSLS estimators. In addition, we develop a sensitivity analysis framework to assess the robustness of the estimated causal effects when key proxy structural conditions are violated. Extensive simulation studies demonstrate that the proposed method performs consistently well across various settings, effectively accounting for unmeasured confounding and capturing bidirectional causality.

Empirical evidence on the relationship between abortion rates and murder rates from U.S. state-level data further illustrates the importance of considering bidirectional relationships in policy-relevant research. By examining this relationship from both directions simultaneously, we detect a robust negative effect of abortion rates on murder rates, consistent with previous literature, while also detecting a 
potential effect from murder rates to abortion rates that conventional unidirectional research failed to identify. The sensitivity analysis reveals substantial differences between the two causal directions. While the abortion rates to murder rates effect maintains a consistent negative direction across different model specifications, the reciprocal effect exhibits greater sensitivity to the proxy structural conditions. This finding exemplifies the importance of considering the sensitivity analysis to assumption violations for causal effect conclusions.

The proposed methods can be improved and extended in several directions. First, we consider identification and estimation within the linear proximal model framework. Future research could explore bidirectional causal relationships using more flexible approaches, such as additive models or nonparametric methods. Additionally, while our estimation methods focus on continuous outcome variables, investigating identification and estimation for binary outcome variables would also be valuable in practice \citep{LiuAJE2024}. The study of these issues is beyond the scope of this paper and we leave them as future research topics.

\section{Competing interests}
No competing interest is declared.

\section{Acknowledgments}
The authors would like to thank Professor Wang Miao from Peking University for his valuable comments on the simulation studies and real data analysis.

\section{Author contributions statement}
Jiaqi Min and Shanshan Luo conceived the experiment, conducted the experiment and analyzed the results.  Jiaqi Min, Xueyue Zhang, and Shanshan Luo wrote and reviewed the final manuscript.

\section{Data availability}
All simulation and real data analysis code to reproduce this work are available at Github \href{https://github.com/Jiaqi-Min/Bidirectional-Proximal-Causal-Inference}{Jiaqi-Min/Bidirectional-Proximal-Causal-Inference}. The real-world data for this paper are not publicly available because of licensing restrictions.

\bibliographystyle{abbrvnat}
\bibliography{reference}
\end{document}